\documentclass[espf]{aastex}
\usepackage{emulateapj5}

\slugcomment{Printed \today}

\shorttitle{Observations and Modeling of MIR Emission from HMPOs}
\shortauthors{De Buizer, Osorio, and Calvet}

\begin{document}

\title{Observations and Modeling of the 2 -- 25 $\micron$ Emission From High
Mass Protostellar Object Candidates}

\author{James M. De Buizer}
\affil{Gemini Observatory, Casilla 603, La Serena, Chile;
jdebuizer@gemini.edu  }

\author{Mayra Osorio}
\affil{Instituto de Astrof\'\i sica de Andaluc\'\i a, CSIC,
Camino Bajo de Hu\'etor 50, E-18008 Granada, Spain; osorio@iaa.es}

\and

\author{Nuria Calvet}
\affil{Harvard-Smithsonian Center for Astrophysics,
60 Garden Street, Cambridge, MA 02138, USA; ncalvet@cfa.harvard.edu}

%\altaffiltext{1}{E-mail: jdebuizer@gemini.edu}

\begin{abstract}

This is a report on detailed modeling of young high-mass protostellar
candidates during their most embedded and obscured phases. We performed
narrowband mid-infrared imaging of three candidate high-mass protostellar
objects in G11.94-0.62, G29.96-0.02, and G45.07+0.13 at Gemini
Observatory using the Thermal-Region Camera and Spectrograph (T-ReCS). The
sources were imaged through up to 11 narrowband filters, sampling their SEDs
over the entire 2--25 $\mu$m infrared range. For the first
time, we have fit the observed SEDs of  massive protostars with
models that take into account
departures from spherical symmetry in the infalling envelopes. In this way, we
have been able to back out of the models detailed physical parameters for these
earliest stages of massive stellar life. Our detailed modeling suggests that massive star formation can
proceed in a way very similar to the formation of low-mass stars.

\end{abstract}


\section{Introduction}

The most massive stars begin their lives in the darkest recesses of giant
molecular clouds, obscured from our view by a thick veil of dust and gas.
Most of the time the earliest stages of massive star formation cannot be
observed at visual wavelengths, and in many cases even the near-infrared
emission from these protostars cannot escape. We owe much of our ignorance
of the massive star formation process to the fact that we have a very
limited range of wavelengths at our disposal with which to observe these
youngest stages of massive stellar evolution.

The problem is further compounded by the fact that massive OB stars
($\geq$ 8 M$_{\sun}$) are relatively few in number and, like all stages of
massive stellar evolution, the formation process proceeds very quickly. It
is therefore difficult to catch massive stars in the act of forming.
Additionally, there is the problem of observing these objects with
sufficient angular resolution. Massive star-forming regions generally lie
at kiloparsec distances, making them more than an order of magnitude
farther away than the closest low-mass star-forming regions. Furthermore,
the percentage of binaries and multiple star systems increases as a
function of mass (Larson 2001), and consequently it is believed that the
vast majority of massive stars do not form alone. It is this combined
distance and multiplicity problem that makes it difficult to resolve and
study in detail a single massive star in its earliest years of formation.

Observations that have been extremely helpful in the study of massive
stars have been cm-wavelength radio continuum and molecular line emission
observations. Interferometric arrays can observe with subarcsecond angular
resolution the radio continuum emission from the photoionized region in
the near-stellar environment around young massive stars. These compact or
ultra-compact \ion{H}{2} (UC \ion{H}{2}) regions are thought to exist
$\sim$10$^5$ years after the formation of a massive star (De Pree,
Rodr\'\i guez, \& Goss 1995). Until recently, this was the earliest
observed phase of massive star formation.

In order to understand better their formation process, we need to observe
massive stars at ages even younger than 10$^5$ years. This means trying to
observe sources before they develop UC \ion{H}{2} regions, which in most
of the cases limits our window of continuum observations to a
relatively narrow
wavelength regime from the mid-infrared ($\sim$3 $\mu$m) to the millimeter
($\sim$1 mm). Observing sources in this ``high mass protostellar object''
(HMPO) phase presents a challenge since the best resolutions achievable
throughout most of this wavelength range (namely the far-infrared to the
sub-millimeter) have, until recently, been on the order of a few to tens
of arcseconds.  However, the bright molecular line emission from these
sources would still be observable with high angular resolution radio
interferometers.

Hot cores of compact and dense molecular gas have been known to exist in
massive star forming regions for several decades now (e.g., Wilson,
Downes, \& Bieging 1979). Some of these ``hot molecular cores'' (HMCs) are
simply externally heated knots of material, but others may have stars forming
at their centers. Recently some HMCs have been
gaining serious attention as possibly forming extremely young massive
stars at their centers (Cesaroni et al. 1994; Kurtz et al. 2000).
Walmsley (1995) proposed that these types of HMCs are undergoing an intense
accretion phase which prevents the development of a UC \ion{H}{2} region.
Since these types of HMCs do not possess radio continuum emission and are
internally heated by massive protostars, they fit into the category of HMPOs.

Though the compositions of these sources are beginning to be understood in
detail through the molecular line studies (e.g., Rodgers \& Charnley
2001), continuum observations remain few because of the angular resolution
problem. However, continuum observations from the mid-infrared to
millimeter have the potential for revealing valuable information about the
massive star formation process. Models of the spectral energy distribution
(SED) of HMCs have been developed under the assumption that they have a
central massive star undergoing spherical accretion of a free-falling
envelope of dust and gas (Osorio, Lizano, \& D'Alessio 1999). The success
of these models in fitting the observable characteristics of HMCs gave
strong support to the hypothesis that some of these objects were true
HMPOs (i.e., not starless HMCs) in a stage previous to the development of
an UC \ion{H}{2} region. However, for that work, only mid-infrared upper
limits were available in most of the sources.  Since the mid-infrared
range of the SED is highly sensitive to the geometry of the source (i.e.,
degree of flattening), in principle, these kinds of observations would
allow testing of more sophisticated models, with a degree of detail
similar to that of those that have been developed for low-mass protostars
(e.g., Osorio et al. 2003).

Recently, observations by De Buizer (2004) and De Buizer et al. (2003)
have identified four candidate HMPOs that have been observed in the
mid-infrared with high-angular resolution. These HMPO candidates lie in
areas of high mass star formation, as evidenced by nearby UC \ion{H}{2}
regions. In fact in two of these four cases, high spatial resolution
observations from 8m telescopes were needed to simply isolate the
mid-infrared emission of the HMPOs from the nearby UC \ion{H}{2} regions.
This means that, though more sensitive than what is achievable from the
ground, data from infrared satellites like \emph{MSX} ($\sim$5$\arcsec$ at
8 $\micron$, $\sim$13$\arcsec$ at 21 $\micron$), and even the
\emph{Spitzer Space Telescope} ($\sim$2$\arcsec$ at 8 $\micron$,
$\sim$6$\arcsec$ at 24 $\micron$), are in general not adequate for these
studies because of their relatively poor angular resolutions. By being
able to isolate emission from the HMPO alone, mid-infrared observations
from ground-based 8 to 10m-class telescopes represent a invaluable tool in
the study of these earliest stages of massive star formation.

In this article we present new sub-arcsecond mid-infrared observations of
three of the four candidate HMPOs identified by De Buizer (2004)
and De Buizer et
al. (2003). The fourth HMPO candidate, G305.20+0.21, will be the
subject of a following article. Using the Gemini South telescope we
imaged these three HMPO
candidates through up to 11 filters, sampling their SEDs over the entire 2--25
$\mu$m infrared atmospheric window. Our motivation for performing
these observations was to sample the SEDs of HMPOs with sufficient
spectral resolution and coverage to accurately test new SED models. These
models use envelopes that take into account deviations from spherical
geometry in their inner region due to rotation and at large scales due to
natural elongation of the source.  By matching these models to the data we
attempt to ``back out'' of the models detailed physical parameters for
these earliest stages of massive stellar life. In this way, we hope to be
able to learn some details about the massive star formation process during
these highly obscured stages, as well as the strength and limitations of
the non-spherical SED models.

\section{Observations}

Observations were carried out at Gemini South over the time period between
2003 September and 2004 November. Imaging was performed with the
Thermal-Region Camera and Spectrograph (T-ReCS). The instrument employs a
Raytheon 320$\times$240 pixel Si:As IBC array which is optimized for use
in the 7--26 $\mu$m wavelength range, but can perform with modest
sensitivity in the 2--5 $\mu$m wavelength range as well. The pixel scale
is 0.089$\arcsec$/pixel, yielding a field of view of
28$\farcs$8$\times$21$\farcs$6. Sky and telescope subtraction were
achieved through the standard chop-nod technique.

Images were performed with \emph{K}, \emph{L}, and \emph{M} filters,
as well as six narrowband
silicate filters encompassing the full 10 $\mu$m \emph{N}-band spectral window,
and two narrowband filters within the 20 $\mu$m \emph{Q}-band spectral window.
Effective central wavelengths for these filters and their spectral range
are given in Table 1.  For G11.94-0.62 and G45.07+0.13 we also have images
from De Buizer et al. (2003) through the \emph{Q3}
($\lambda_o$=20.81 $\mu$m, $\Delta\lambda$=19.99--21.64 $\mu$m) filter
from the NASA Infrared Telescope Facility (IRTF) using the instrument
MIRLIN.

Table 1 also lists the on-source integration times through each of the
T-ReCS filters for each HMPO candidate field. In addition, this table
lists the standard star used to flux calibrate each image, and the assumed
flux density for that standard through the given filter. These assumed
standard star flux densities were found by convolving the spectral
irradiance templates of the stars from Cohen et al. (1999) with the given
T-ReCS filter transmission profile.

Flux densities were derived for each filter for the HMPO candidate as well
as any other mid-infrared sources on the field. Tables 2-4 list the flux
densities for all sources. These flux densities are quoted with their
1-$\sigma$ total error, which is a quadrature addition of the statistical
variation from the aperture photometry (due to the standard deviation of
the background array noise) and the flux calibration error. Usually the
flux calibration error is found from the variation of the standard star
flux throughout the course of the night, however our observations
were taken piece-wise on many nights throughout several semesters.
Flux calibration was therefore achieved by
observing a standard star at a similar airmass just before or after the
observation of the target science field. Therefore, we do not have
calibration variability statistics for each filter on each night the data
were taken. However, the calibration factor [ratio of accepted flux in
Janskys to analog-to-digital converter units (ADUs) per second per pixel]
derived from the standard star observations through each filter varied
little throughout 14 months of data collection, and so we use here the
standard deviation of the calibration factor over this 14 month period as
the flux calibration error. This can be considered an extremely
conservative estimate of the errors, taking into account a wide variety of
atmospheric and observing conditions that affect flux calibration. Most
filters have modest flux calibration errors (for mid-infrared
observations) with standard deviations between 2 and 10\%. Flux
calibration through certain filters is more difficult due to the presence
of various atmospheric absorption lines contaminating the filter bandpass,
some of which can be highly variable. Those filters most affected are the
7.7 $\mu$m (21\%), 12.3 $\mu$m (19\%), 18.3 $\mu$m (15\%), and 24.6 $\mu$m
(23\%) filters.

Furthermore, the continuum observed through these filters may be affected
by the presence of certain spectral features common to young stellar
objects. In particular, the 12.3 $\mu$m continuum flux densities for the
targets may be skewed to higher values due to the presence of bright
[NeII] lines at 12.8 $\mu$m (see for instance, Faison et al. 1998). The
18.3 $\mu$m and 20.8 $\mu$m observations can also be skewed high or low
depending on the presence in either emission or absorption of the 18
$\mu$m silicate feature.

\section{The Targets}

The three sources below were chosen because they are among the only HMPO
candidates known to be bright in the mid-infrared (De Buizer 2004, De
Buizer et al. 2003)  and are sufficiently far enough away from nearby
\ion{H}{2} regions to be studied in detail without confusion. All three
sources are thought to be HMPOs primarily because they are sources found
in regions of known high mass star formation yet have no UC \ion{H}{2}
emission of their own, and all are coincident with molecular
maser emission which is
thought to trace massive star formation and its processes. Each source has
further unique evidence indicating that it is a HMPO, which we discuss
in detail below. A summary of the observations for each object will also
be discussed in this context.

\subsection{G11.94-0.62}
\label{G11m}

Of the four sources identified as mid-infrared bright HMPO candidates by
De Buizer (2004) and De Buizer et al. (2003), the status of the source in
G11.94-0.62:DRT03 1 as a genuine HMPO is the most uncertain. However the HMPO
candidate in G11.94-0.62 satisfies, at the very least, the minimum
criteria outlined at the beginning of this section. DRT03 1 was
found as an unresolved mid-infrared source in the survey of De Buizer et
al. (2003) located at the site of a water maser clump offset
$\sim$10$\arcsec$ from a cometary UC \ion{H}{2} region. DRT03 1 has no
detected cm radio continuum emission of its own.

However, G11.94-0.62 has not been observed with high spatial resolution
molecular line imaging. Some molecular tracers have been discovered toward
the region, like CS (Bronfman et al. 1996) and NH$_3$ (Cesaroni, Walmsley,
\& Churchwell 1992), but it is not known if this emission is related to
the nearby UC \ion{H}{2} region, the HMPO candidate, or both. While it is
encouraging that some molecules have been found in the region, CH$_3$CN (a
tracer of denser and hotter molecular material and indicator of the
presence of HMCs) has been searched for in the region and was not detected
(Watt \& Mundy 1999). The UC \ion{H}{2} region was also imaged at 2.7 mm
by Watt \& Mundy (1999), but there was no detection at the site of DRT03 1.

De Buizer et al. (2003) derive a lower limit to the bolometric luminosity
for DRT03 1, finding that it is at least as
bright as a B9 star. Though this spectral type is derived from an extreme
lower limit to the bolometric luminosity, it does not appear that this
source is a massive (M$>$10$M_{\odot}$) protostar, but more
likely an intermediate mass protostellar object. If this is true, given
its observational similarities to its higher mass counterparts, it will be
interesting to study the similarities and differences in detail to learn
more about the interrelation of the intermediate and high mass star
formation processes. It would also have interesting consequences for the
hypotheses regarding excitation of water masers by lower mass
(non-ionizing) stars (Forster \& Caswell 2000).

Our sensitive Gemini mid-infrared observations of the region around
G11.94-0.62 have revealed extended dust structure, the likes of which are
not seen at any other spectral regime (Figure 1). These images at the
sub-arcsecond angular resolution of Gemini confirm the results of the
lower resolution observations of De Buizer et al. (2003) that the
mid-infrared morphology of the UC \ion{H}{2} region is not cometary, as is
seen in cm radio continuum emission (Wood \& Churchwell 1989). The HMPO
candidate, DRT03 1, is clearly detected at shorter mid-infrared
wavelengths but is only a 3-$\sigma$ and 4-$\sigma$ detection at 20.8 and
24.6 $\mu$m, respectively.

With no information on the Rayleigh-Jeans side of the SED for the HMPO
candidate alone, we will not be able to constrain the models at the longer
wavelengths. Table
2 lists the observed flux densities for the HMPO candidate as well as for
the other mid-infrared sources in the field. It is not clear if each of
these mid-infrared sources house a stellar component or if they are simply
knots of dust in an extensive \ion{H}{2} region.

\subsection{G29.96$-$0.02}
\label{G29dB}

The HMPO candidate in G29.96$-$0.02 is sufficiently well-studied to be
considered a prototype. The hot molecular component of this source was
first discovered in the observations of Cesaroni et al. (1994). In this
work, a hot (T$_K$$>$ 50 K) core of molecular material was detected, traced
by NH$_3$(4,4) line emission near to, but offset $\sim$2$\arcsec$ from a
cometary-shaped \ion{H}{2} region. This hot molecular core was found to be
coincident with a cluster of water masers. Cesaroni et al. (1994)
conjectured that this source, like the other HMCs in their survey, is
internally heated by a developing massive protostar. However, as was
mentioned in the introduction, some HMCs could be
externally heated by the nearby massive stars powering the UC \ion{H}{2}
regions. In the case of G29.96-0.02, the millimeter line study of Gibb,
Wyrowski, \& Mundy (2003) provided strong evidence that the HMC is
internally heated. Therefore, the HMC in G29.96-0.02 appears to be a
genuine HMPO.

The HMPO in G29.96-0.02 was directly imaged for the first time in the
mid-infrared by De Buizer et al. (2002) on the Gemini North 8-m telescope.
The HMPO lies $\sim$3$\arcsec$ away from the radio continuum peak of the
bright \ion{H}{2} region and the emission from the UC \ion{H}{2} region
peak cannot be resolved from the HMPO on 4-m class telescopes (Watson et
al. 2003). However, the large aperture of Gemini has sufficient resolving
power in the mid-infrared ($\sim$0$\farcs$5) to pull in the peak of the
dust emission from the \ion{H}{2} region to allow direct observation of
the HMPO.  Even so, the resolved thermal dust emission from the \ion{H}{2}
region is still extensive, and therefore we cannot avoid observing the
HMPO overlaid on a diffuse background of emission (Figure 2). Following
the technique outlined by De Buizer et al. (2002), one can fit this
background emission and subtract it off, effectively leaving only the
mid-infrared emission of the HMPO. This technique was employed for the
images obtained in the work presented here. In Table 3 we present the flux
densities for the HMPO candidates and the UC \ion{H}{2}, as well as the
flux density of the HMPO candidate without subtracting off the background
due to the extended UC \ion{H}{2} emission. This was done so that it is
transparent to the reader how much flux has been subtracted off, and what
the absolute upper limits are for the flux densities of the HMPO candidate.

The G29.96-0.02 HMPO has been studied in great detail in molecular line
studies (e.g., Cesaroni et al. 1998, Wyrowski, Schilke, \& Walmsley 1999;
Maxia et al. 2001.) However, other than the mid-infrared observations of
De Buizer et al. (2002) and those presented here, only a few continuum
observations at other wavelengths exist. This source has been observed at
1 and 3 mm by Maxia et al. (2001). However, as pointed out by
De Buizer et al. (2002), there appears to be problems with these
flux densities since
they lead to a negative value (see Figure \ref{fig6})  for the index of
dust opacity, $\beta$, whereas most models assume 1$\leq$$\beta$$\leq$2
for $\lambda$$\geq$200 $\mu$m.

Another observation at 3 mm was carried out by Olmi el al. (2003). They
did not resolve the HMPO, and therefore there might be possible
contamination from the nearby UC \ion{H}{2} region. They try to obtain the
flux density of the HMPO at 3 mm, making a subtraction of the free-free
contamination estimated from the 2 cm image. This measurement together
with the detection of the HMPO at 1.4 mm carried out by Wyrowski et al.
(2005, personal communication)  indicate that G29.96-0.02 HMPO is a strong
emitter at millimeter wavelengths.

\subsection{G45.07+0.13}

Listed as a ``known hot core'' in the review article of Kurtz et al.
(2000), the HMPO in G45.07+0.13 gained attention mostly through the work
of Hunter, Phillips, \& Menten (1997). However most of the observations by
Hunter, Phillips, \& Menten (1997) are low spatial resolution continuum
and molecular line maps observed towards UC \ion{H}{2} regions. In the
case of G45.07+0.13 they found several molecular tracers, some believed to
be in outflow, emanating from the location of the UC \ion{H}{2} region.
Hunter, Phillips, \& Menten (1997) contest that there is a single star
here, embedded in a molecular core, at a very early stage of development
such that it has just begun to develop a UC \ion{H}{2} region. For these
reasons it has been labeled a hot core.

The observations of De Buizer et al. (2003) showed that there are three
mid-infrared sources at this location, all within 8$\arcsec$ of each
other. The brightest mid-infrared source on the field is coincident with
the UC \ion{H}{2} region. However, at a location $\sim$2$\arcsec$ north of
the UC \ion{H}{2} region there is a group of water masers associated with
a mid-infrared source that has no cm radio continuum emission (Figure 3).
De Buizer et al. (2003) showed that this northern mid-infrared source,
DRT03 3, has
a very bright mid-infrared luminosity of $\sim$4$\times$10$^3$
L$_{\odot}$. Assuming that the mid-infrared luminosity is equal to the
bolometric luminosity (a gross underestimate), DRT03 3 has the
equivalent luminosity of a B0 star. We therefore have a source luminous
enough to be a massive protostar, situated at the location of a group of
water masers, yet this source has no cm radio continuum emission of its
own. For these reasons, De Buizer et al. (2003) claim that DRT03 3 is
the true HMPO in this field\footnote{The coordinates for the HMPO
candidate are wrong in De Buizer et al. (2003). See the caption of Figure
3 for correct coordinates}.

Although there are multiple sources of thermal dust emission present here,
it does appear from the high spatial resolution CS maps of Hunter,
Phillips, and Menten (1997) that the stellar source powering the UC
\ion{H}{2} region is the only viable source of the outflow in the region
(contrary to what was stated by De Buizer et al. 2003). Since no such
tracers are seen coming from the HMPO candidate, this perhaps confirms the
youth of the object, placing it in a stage of formation before the onset
of an outflow. While the resolution of the 800 and 450 $\mu$m maps of
Hunter, Phillips, \& Menten (1997) are too coarse to ascertain which
mid-infrared source may be dominating the emission at these longer
wavelengths, their higher resolution ($\sim$2$\arcsec$) 3 mm map shows no
emission from the location of DRT03 3, though emission is
detected towards the other two mid-infrared sources on the field.

Our new observations of this field with Gemini have revealed the presence
of four more faint mid-infrared sources present (Figure 3) in addition to
those already found by De Buizer et al. (2003). It appears that this may
be a small cluster of star formation centered on the O9 star powering the
UC \ion{H}{2} region. DRT03 3 is seen at all 11 mid-infrared
wavelengths, as is the UC \ion{H}{2} region (DRT03 2). Table 4 lists the flux
densities for all mid-infrared sources on the field. Unfortunately, we
have no detections of the HMPO candidate alone at any wavelength on the
Rayleigh-Jeans side of the SED, and have to rely upon integrated flux
density measurements from the whole region (Su et al. 2004) as upper
limits to constrain the models.

\section{Modeling}

\subsection{Description of the Model}

We will calculate the SED of a HMPO by modeling it as an envelope of gas
and dust that is freely falling onto a recently formed massive central
star, which is responsible for the heating.  Osorio, Lizano \& D'Alessio
(1999) modeled the SED of a sample of such objects by assuming that the
envelope has spherical symmetry. For that work, only upper limits were
available for the mid-infrared flux densities of most of the sources.
Because this wavelength range is very sensitive to the geometry of the
source, here we extend that work by including rotation and flattening of the
envelope in the models, so that the results can be tested with the new
mid-infrared data presented in this paper. For this purpose, we have followed
procedures similar to those already developed in low-mass star models
(Osorio et al. 2003).

Envelopes with a density distribution such as that given by Terebey, Shu,
\& Cassen (1984, hereafter TSC envelopes) assume that the rotation of the
infalling material becomes important only in the vicinity of the
centrifugal radius, $R_c=r_0^4 \Omega_0^2/GM_*$, where $\Omega_0$ is the
angular velocity (assumed to be constant and small) at a distant reference
radius $r_0$, and $M_*$ is the mass of the central protostar.

The TSC envelopes are flattened only in the inner region, being
essentially spherical in the outer region. The inner regions ($r<r_0$) of
these envelopes are described by the solution of Cassen \& Moosman
(1981) and Ulrich (1976) (hereafter referred as CMU). The solution is
determined by specifying the velocities and densities at the (distant)
reference radius $r_0$, where the velocity is nearly equal to the radial
free-fall value, with only a small azimuthal component corresponding to a
constant angular velocity, and the density is set by the required mass
infall rate $\dot M$. With these assumptions, the density of the inner
envelope ($r<r_0$) is given by
 $$\rho_{\rm CMU}(r,\theta)={{\dot M}\over{4\pi(GM_*r^3)^{1/2}}}$$
 $$\times {\left(1 + {\cos\theta \over \cos\theta_o} \right)^{-1/2}}
{\left({\cos\theta \over \cos\theta_0} + {2 R_c \cos^2\theta_0 \over
r} \right)^{-1}},$$
 where $\theta$ and $\theta_0$ are the position angles with respect to the
rotational symmetry axis of the system at radial distances from the
central mass $r$ and $r_0$, respectively. At large distances ($r\gg R_c$)
the motions are radial ($\theta \rightarrow \theta_0$) and the density
tends to the density distribution for radial free-fall at constant mass
infall rate. For $r \la R_c$, the motions become significantly non-radial,
and the angular momentum of the infalling material causes it to land on a
disk ($\theta=90^\circ$) at distances $0\le r \le R_c$. Therefore, $R_c$
is the largest radius on the equatorial plane that receives the infalling
material.

A more complex description can be given in terms of intrinsically
flattened envelopes, such as those with the density distribution resulting
from the gravitational collapse of a sheet initially in hydrostatic
equilibrium (Hartmann et al. 1994; Hartmann, Calvet, \& Boss 1996). These
envelopes are flattened not only in their inner region, because of
rotation (as the TSC envelopes), but also at large scales due to the
natural elongation of the cloud. If the rotational axis is perpendicular
to the plane of the self-gravitating layer, the density distribution of
the infalling material will be axisymmetric and can be written as:
 $$\rho(r,\theta)=\rho_{\rm CMU}(r,\theta) \eta \cosh^{-2}(\eta
\cos\theta_0) \tanh^{-1}(\eta),$$
 where $\rho_{\rm CMU}$ is the CMU density distribution for the same $\dot
M$ and $R_c$, and $\eta$ is a measure of the degree of flattening
(Hartmann et al. 1996). The degree of flattening in these envelopes is
measured by the parameter $\eta\equiv R_{\rm out}/H$, where $R_{\rm out}$
is the outer radius of the envelope and $H$ is the scale height.
Hereafter, we will designate them as $\eta$-envelopes.

The scale of this density distribution can be set by using the parameter
$\rho_1$, defined as
 $$\rho_1\equiv {{\dot M}\over{4\pi (2G M_* r_1^3)^{1/2}}},$$
 where $r_1$ is a reference radius. This reference density corresponds to
the density that a spherically symmetric free-falling envelope with the
same mass accretion rate, $\dot M$, would have at the reference radius
$r_1$. In our modeling, we will use $r_1=1$ AU, and we will designate the
reference density as $\rho_{\rm 1\,AU}$.

The remaining parameters that define the density distribution of the
$\eta$-envelopes are the centrifugal radius, $R_c$, and the inclination of
the polar (rotational) axis to the line of sight, $i$. Since
$\eta$-envelopes are flattened not only in their inner region but also at
large scales most of their material is accumulated over the equatorial
plane, having an extinction lower than a TSC envelope of the same mass
provided the line of sight is not oriented too close to the edge-on
position. Therefore, a flattened $\eta$-envelope model predicts stronger
near and mid-infrared emission and less deep absorption features than
those predicted by a TSC envelope model of the same mass. Given that our
sources are strong mid-infrared emitters, we will preferentially use the
$\eta$-envelope density distribution to model our sources.

The temperature distribution along the envelope, as well as the SED from 2
$\mu$m to 2 cm, are determined following the procedures described by
Kenyon, Calvet, \& Hartmann (1993, hereafter KCH93), Calvet et al. (1994),
Hartmann et al.  (1996), and Osorio et al. (2003).  For a given luminosity
of the central star, $L_*$, which is assumed to be the only heating source
(i.e., neglecting other possible sources of heating such as accretion
energy from a disk or the envelope), the temperature structure in the
envelope is calculated from the condition of radiative equilibrium, using
the approximation of an angle-averaged density distribution. Nevertheless,
the emergent flux density is calculated using the exact density
distribution. Because we want to compare our results with high spectral
resolution mid-infrared data, our model spectra wavelengths were finely
sampled ($\Delta$log$\lambda$ = 0.02) in the range
3-25~$\mu$m in order to define with high spectral resolution the 10~$\mu$m
silicate absorption feature. Outside this range, wavelengths were sampled
at $\Delta$log$\lambda$ = 0.2-0.3. Additionally, in order to make a
proper comparison with the observations, we convolved our results with the
T-ReCS filter transmission profile. Given that the spectral response of
the filters is fairly constant, we do not find large variations (only
$\sim$5\%) between the averaged and original spectrum.

In our models we use a recently improved dust opacity law, corresponding
to a mixture of compounds that was adjusted by matching the observations
of the well known low-mass protostar L1551 IRS5 (Osorio et al. 2003). This
mixture includes graphite, astronomical silicates, troilite and water ice,
with the standard grain-size distribution of the interstellar medium,
$n(a)\propto a^{-3.5}$, with a minimum size of 0.005~$\mu$m and a maximum
size of 0.3~$\mu$m. The details of the dust opacity calculations are
described by D'Alessio, Calvet, \& Hartmann (2001).

The inner radius of the envelope, $R_{\rm in}$, is taken to be the dust
destruction radius, that is assumed to occur at a temperature of 1200 K,
corresponding to the sublimation temperature of silicates at low densities
(D'Alessio 1996).  The centrifugal radius is assumed to be considerably
larger than the dust destruction radius in order to have a significant
degree of flattening due to rotation in the inner envelope. The outer
radius of the envelope, $R_{\rm out}$, is obtained from our 11.7~$\mu$m
images.

In summary, once the dust opacity law and the inner and outer radius of
the envelope are defined, the free parameters of our model are $\eta$,
$\rho_{\rm 1\,AU}$, $R_c$, $i$, and $L_*$. In the following, we will
discuss on the contribution of these parameters to the resulting SED.

\subsection{Behavior of the SED}

In order to illustrate the effect of the flattening of the envelope, we
have calculated the resulting SED for a TSC envelope and compared it with
an $\eta=2$ envelope of the same mass. The results are shown in Figure 4,
where it can be seen that the flattened $\eta=2$ envelope predicts
stronger near and mid-infrared emission, and less deep absorption features
than the TSC envelope, both for low ($i=30^\circ$) and moderately high
($i=60^\circ$) inclination angles.  This behavior is a consequence of the
flatter distribution of material in the $\eta=2$ envelope, resulting in an
extinction lower than the TSC envelope over a wide range of viewing angles
(except for a nearly edge-on view, $i=90^\circ$, where the behavior would be
the opposite). Thus, in general, the flattened $\eta$-envelopes allow the
escape of a larger amount of near and mid-infrared radiation, and consequently
they predict stronger flux densities than the TSC envelopes in this
wavelength range. Since our sources are strong infrared emitters, in our models
we will use $\eta$-envelopes with a significant degree of flattening,
$\eta$, of the order of 2.

The observed SEDs of our sources are well sampled in the near and
mid-infrared wavelength range. However, this wavelength coverage in
insufficient for the models to determine the value of the bolometric
luminosity (that we take to coincide with the stellar luminosity, $L_*$).
The bolometric luminosity is well constrained by the far-infrared data
(where the peak of the SED occurs), and to a less extent by the
submillimeter or millimeter data.  Unfortunately, the available
far-infrared data for our sources do not have the appropriate angular
resolution to avoid contamination from other nearby sources, and should
be considered as upper limits, while millimeter data are only available
for one of our sources. Therefore, in general, we constrain the
value of the luminosity by fitting the mid-infrared data closer to the SED
peak, that are helpful to set a lower limit for the luminosity.

On the other hand, the depth and shape of the silicate absorption feature
(which is very well delineated by our mid-infrared data) determines the
extinction along the line-of-sight, that in turn is related to the
inclination angle, $i$, the density scale, $\rho_1$, and the centrifugal
radius, $R_c$. The depth of the 10 $\mu$m silicate absorption increases
not only with the density scale of the envelope, but also with the
inclination angle, since the extinction is much smaller along the
rotational pole ($i=0$) than along the equator ($i=90^\circ$) (see Osorio
et al.  2003).  Thus, the behavior of the parameters $\rho_1$ and $i$ is
somewhat interchangeable in the mid-infrared (although such behavior is
different at millimeter wavelengths). For instance, an envelope with a
combination of low density and high inclination could have an extinction
and mid-infrared emission similar to an envelope with a higher density but
viewed pole-on. Therefore, for the sources where millimeter data are not
available, it is difficult to constrain $i$ and $\rho_1$ simultaneously, and
we will search for fits both for high as well as for low inclination
angles.

The behavior of the SED with the centrifugal radius, $R_c$, is more
complex. Increasing the value of $R_c$ will widen the region of evacuated
material, producing an overall decrease of the extinction along the line
of sight, but, at the same time, it will result in a decrease of the
temperature of the envelope, because the material tends to pile up at
larger distances from the source of luminosity. Thus, depending on the
wavelength range, one of the two effects will be the dominant. At short
wavelengths, where the dust opacity is higher, it is expected that the
lower extinction of the envelopes with large values of $R_c$ will result
in an increase of the observed emission. At longer wavelengths, where the
opacity is low enough (except in the silicate absorption feature), the
dominant effect is expected to be due to the decrease in the temperature,
resulting in a decrease of the observed emission for larger values of
$R_c$.  This is illustrated in Figure 4, where it is shown that the
emission of the envelope with $R_c=600$ AU dominates in the near-infrared
range, while the emission of the $R_c=250$ AU envelope dominates at longer
wavelengths. As shown in the figure, this result is valid both for low and
relatively high inclination angles, although we note that the point where
the small $R_c$ envelope becomes dominant shifts slightly to longer
wavelengths when increasing the inclination angle (from $\sim 2~\mu$m for
$i=30^\circ$ to $\sim 5~\mu$m for $i=60^\circ$). This shift occurs because
envelopes seen at higher inclination angles are more opaque, thus
requiring a longer wavelength (where the opacity is lower) for the small
$R_c$ envelope to become dominant.

Since we expect the value of $R_c$ to be of the order of the radius of a
possible circumstellar disk, in our modeling we will explore a range of
values for $R_c$ from tens to hundreds of AUs, depending on the luminosity
of the source. The upper limit of this range is suggested by the highest
angular resolution observations of disks reported towards massive
protostars (e.g., Shepherd, Claussen, \& Kurtz 2001), while the lower
limit is set by the smallest values recently reported for protoplanetary
disks around low-mass YSOs (Rodr\'\i guez et al. 2005).

 From the values of the parameters obtained from our fits to the observed
SEDs we can derive other physical parameters, such as the total mass of
the envelope, $M_{\rm env}$, obtained by integration of the density
distribution, and the mass of the central star, $M_*$, derived from its
luminosity, $L_*$. Once the mass of the star is known, the mass accretion
rate in the envelope, $\dot M$, can be derived from the value of $\rho_1$.

Yorke (1984) and Walmsley (1995) proposed that some objects could be
accreting material at a rate high enough to quench the development of an
incipient HII region. These objects would lack detectable free-free
emission, being good candidates to be the precursors of the UCHII regions.
In the case of spherical accretion, it is easy to show that the critical
mass accretion rate required to reduce the HII region to a small volume
close to the stellar surface is $\dot M_{\rm crit}=(8 \pi G m^2_{\rm H}
\alpha^{-1} \dot N_i M_*)^{1/2}$, where $m_{\rm H}$ is the hydrogen mass,
and $\alpha$ is the recombination coefficient (excluding captures to the
$n=1$ level). We will adopt the value, $M_{\rm crit}$, as a critical measure of the mass accretion rate required to ``choke off'' the possible development of an UC \ion{H}{2}
region. Since we are assuming departures from spherical symmetry in our envelopes, in reality a more complex calculation would be required to
accurately derive the corresponding critical mass rate. However, for simplicity we will assume that if the ratio between the mass accretion rate in our envelope and the critical spherical mass accretion rate $\dot M /\dot M_{\rm crit} \gg 1$, it would indicate that photoionized emission is not expected to be detectable in the source.

\section{Comparison of the Models with the Observations}
\label{models}

In this section, we estimate the physical parameters of the three HMPO
candidates by fitting our models to the observed SEDs, using also the
available information on size and morphology inferred from the images.
Given that spherical envelopes do not appear to be able to account for
the observed properties of the SEDs in the mid-infrared, we have modeled the
sources using $\eta$-envelopes with a significant degree of elongation
(see discussion in $\S$4.2). Since the temperature distribution in the
envelopes is estimated using the approximation of an angle-averaged
density distribution, valid for moderately elongated envelopes, we did
not attempt to consider very high values of $\eta$, that correspond to
extremely elongated envelopes where the temperature calculations probably
would not be accurate enough. Therefore, we explored values of $\eta$
around 2, and in all three sources we were able to get good fits for
$\eta=2.5$. We note that this value coincides with the one that provided
the best fit for the low-mass protostar L1551-IRS5, after exploring a
wide range of values of the $\eta$ parameter (Osorio et al. 2003).

Given the incomplete and non-uniform coverage of the observed SED, the
fitting process cannot be automated and should be carried out manually on
a case-by-case basis. As was already mentioned in the previous section,
the luminosity is well constrained by the peak of the SED, that is
expected to happen in the far-infrared. Unfortunately, at present, far-infrared
observations do not reach the angular resolution required to properly
isolate the emission of these distant sources from contamination of other
sources in their vicinity. Therefore, in general, the far-infrared data are
considered only as upper limits and we have to use the mid-infrared data to
obtain a range of possible values for the luminosity. Thus, the general
strategy was to first run a set of models with increasing luminosities
until we found a value that was able to reproduce the observed mid-infrared flux
density near the peak of the SED and consistent with the millimeter
and submillimeter data.  Once an approximate value of the luminosity was
found, we proceeded to run a set of models to find the density scale, which
is then basically determined by the millimeter and submillimeter data
points. When the density scale was fixed, we ran models with different
inclinations in order to find the best fit to the absorption feature in
the mid-infrared, which is particularly sensitive to the value of the
inclination angle. Finally, we refined the modeling by testing a range of
possible values of the centrifugal radius.

If the data point coverage of the SED is good enough we should end this
process with essentially one final model (the ``best fit model'') that is
able to reproduce the observed data points within the constrains imposed
by the observational uncertainties. In the case that only upper limits are
available in the millimeter and submillimeter domain, the density scale
and inclination angle cannot be constrained simultaneously, and in this
case we present two fits, one for a low inclination angle, and a second
one for a high inclination angle. The value of the centrifugal
radius not well constrained in this case either. However, on the basis of an
additional analysis of the quality of the data points and the
peculiarities of each source (see discussion of individual sources) we
attempt to favor one of the fits as the best fit model.

Table \ref{tbl5} lists the values of the parameters of our best fit
models: the inner, outer and centrifugal radii of the envelope, the
stellar luminosity, the reference density at 1 AU, and the inclination
angle. Our best-fit model SEDs are shown in Figures \ref{fig5},
\ref{fig6}, and \ref{fig7}. Table \ref{tbl6} lists other derived
parameters for each source:  the mass of the envelope, its temperature at
a radius of 1000~AU, the mass and spectral type of the central protostar,
the mass accretion rate, the rate of ionizing photons, and the critical
mass accretion rate. These parameters have been derived following the
procedures explained in the footnotes to the table.

\subsection{G11.94-0.62 HMPO}

This HMPO has been observed through continuum measurements at millimeter
(Watt \& Mundy 1999), submillimeter (Walsh et al. 2003), and mid-infrared
wavelengths. The submillimeter data will be used in our modeling only as
upper limits, since they have been obtained with a large beam
($10''$-$18''$) and therefore it is not possible to separate the emission
of the nearby UC \ion{H}{2} region from that of the HMPO. Likewise, the
millimeter flux densities will be used also as upper limits, since the
analysis of Watt \& Mundy (1999) suggests that the millimeter emission
observed towards this source corresponds to free-free emission of the
nearby UC \ion{H}{2} region with a negligible contribution from dust.

\begin{figure}
\epsscale{1.0}
\plotone{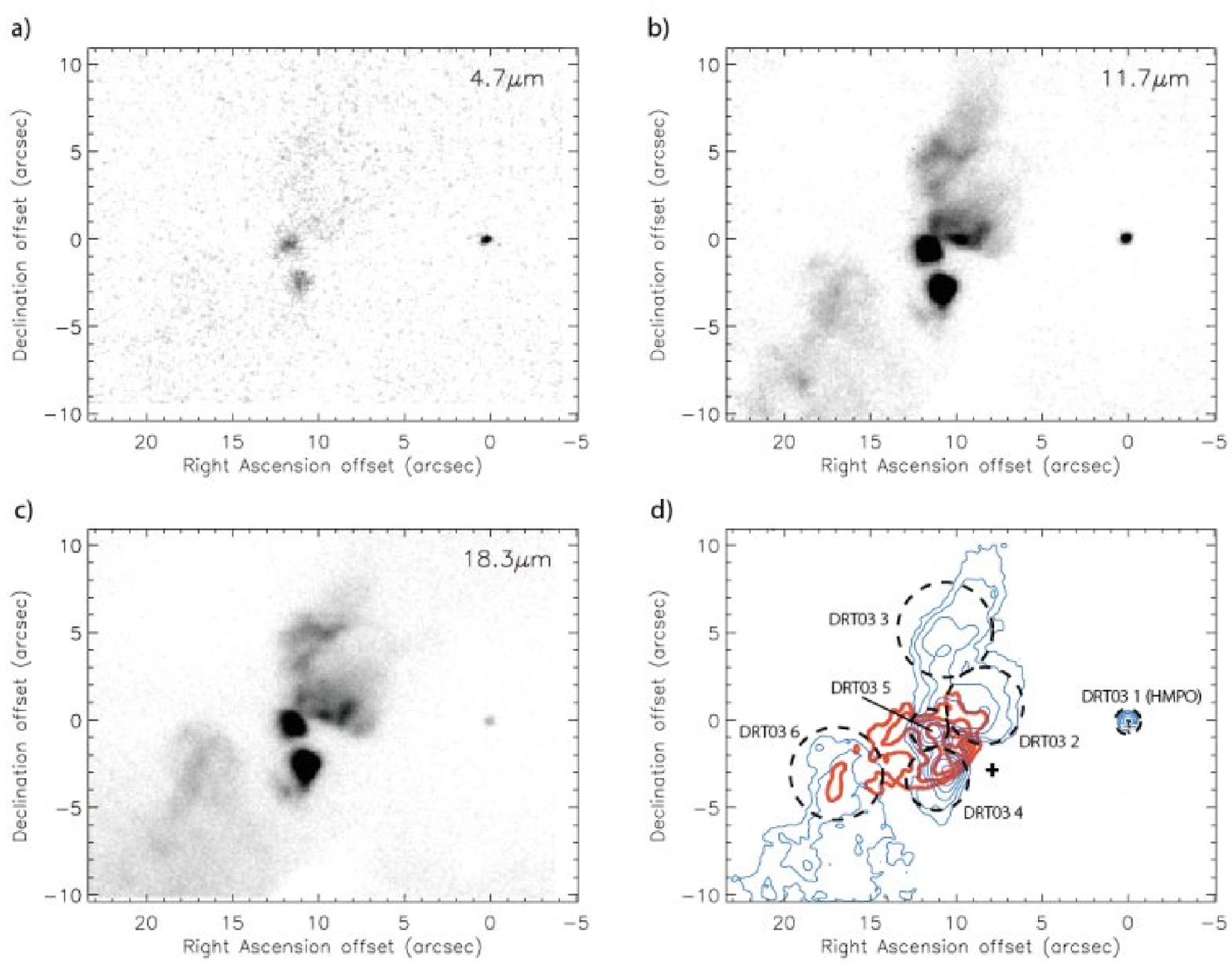}
 \caption{Images of the G11.94-0.62 field taken at a) 4.7 $\mu$m, b) 11.7
$\mu$m, and c) 18.3 $\mu$m at Gemini South with T-ReCS. d) A contour plot
(smoothed by 0$\farcs$36) of the 11.7 $\mu$m image (blue contours) of the
field showing sources with their labels and encircled with dashed lines
that represent the apertures used for photometry. These aperture radii are
given in the notes in Table 2. Plotted as crosses are the locations
of the water maser groups as given by Hofner \& Churchwell (1996).  Also
plotted (red contours) are the 2 cm radio continuum contours from Hofner
\& Churchwell (1996). The origin of each panel is the mid-infrared peak of
the HMPO candidate (R.A.(J2000) = 18$^h$ 14$^m$ $00\rlap.^s29$,
Decl.(J2000) = $-18^\circ$ $53'$ $23\farcs1$).
 \label{fig1}}
\end{figure}

In general, to constrain the physical parameters of the sources, we use
the observed SED as well as the information on size and morphology given
by the images. However, the G11.94$-$0.62 HMPO source appears unresolved in the near and mid-infrared images (see Figure ~\ref{fig1}) and
therefore, we have only an upper limit for its size.

Because we considered the far-infrared, submillimeter and millimeter data
only as upper limits, the luminosity of this source is constrained
essentially by the data at wavelengths near 20~$\mu$m, that indicate that
this is a low luminosity source. We obtain a value of 75 $L_{\odot}$ for
the luminosity of this object, corresponding to a A0 star of 5
$M_{\odot}$. Since we only have upper limits for the flux densities in the
submillimeter-millimeter wavelength range, we cannot constrain the density
and inclination simultaneously. Therefore, we searched for fits both at
low ($i$=20$^{\circ}$-30$^{\circ}$) and high inclination angles
($i$=50$^{\circ}$-60$^{\circ}$). At low inclination angles, we obtained a
quite good fit for $i=$30$^{\circ}$, $\rho_{\rm 1\,AU}=7.5 \times
10^{-13}$ g cm$^{-3}$, and $R_c=30$ AU (dot-dashed line in
Figure~\ref{fig5}), but this model predicts too much emission at 18.3 and
24.6 $\mu$m. In order to reproduce the observational data in the 18-25
$\mu$m wavelength range, we require models with a high inclination angle.
Our best fit model is obtained for a high inclination angle,
$i=53^{\circ}$, with $\rho_{\rm 1\,AU}=$ 2$\times 10^{-13}$ g cm$^{-3}$
($M_{env}=1~M_{\odot}$), and $R_c=30$ AU (solid line in Figure
\ref{fig5}).  We can obtain reasonable fits for $1\times 10^{-13} <
\rho_{\rm 1\,AU} < 4\times 10^{-13}$ g cm$^{-3}$ and $20<R_c<200$ AU. For
example, a model with $R_c=100$ AU is also feasible (dotted line in Figure
\ref{fig5}). However, we favor the $R_c=30$ AU model (solid line in Figure
\ref{fig5}, and Tables 5 and 6) because it reproduces better the flux
densities shortward of 8.7 $\mu$m and the depth of the 10 $\mu$m silicate
absorption feature.

For our favored model, the ratio $\dot M/\dot M_{\rm crit} \gg 1$ (see
Table 6). Therefore, the source could not develop a detectable \ion{H}{2}
region. Furthermore, since the luminosity obtained in this model is quite
low, free-free emission from the HMPO would be hardly detectable, even in
the case of a negligible accretion rate. Thus, G11.94-0.62 is not luminous
enough to be considered as a HMPO.  Most likely, given the low temperature
of the envelope ($T_{\rm 1000\,AU}=40$~K), it would be better considered
as a ``warm core'' that is forming an intermediate-mass star, with an
infall rate more than one order of magnitude higher than the typical
infall rates of low-mass protostars (see KCH93).

\subsection{G29.96-0.02 HMPO}

The G29.96-0.02 HMPO is the only source of our sample that appears to be
associated with strong millimeter emission.  Thus, for this source we have
both infrared and millimeter data to constrain better the models.  The strong
millimeter emission suggests a dense or luminous envelope. In addition,
the relatively strong near-infrared emission observed suggests that
the envelope
should be seen nearly pole-on, since in this direction there is a decrease
of the density with respect to the equatorial plane that allows the escape
of near-infrared radiation.

A reasonable fit (solid-line in Figure \ref{fig6}), explaining
simultaneously the strong millimeter emission as well as the near-infrared
flux densities, is obtained by assuming an extremely dense ($\rho_{\rm
1\,AU}=3.1\times10^{-11}$ g cm$^{-3}$), and therefore very massive
envelope ($M_{env}=576~M_{\odot}$), with a centrifugal radius of $R_c=570$
AU, seen with a low inclination angle ($i=12^{\circ}$), and heated by a
luminous ($L_*=1.8\times 10^4~L_{\odot}$) B1 star (see Tables 5 and 6).
This value of the luminosity is a lower limit, since it is the smallest
value that is consistent with the mid-infrared and millimeter data points, and
we estimate that it can be higher up to a factor of three. Likewise, we
estimate that the uncertainty in the density scale is $\sim$30\%, and that
the range of plausible values is $10^{\circ}<i<20^{\circ}$ for the
inclination angle, and $500<R_c<1200$ AU for the centrifugal radius.
Note that the 9.7 $\mu$m data point is only an upper limit, and therefore
it does not set the depth of the silicate absorption feature. Also, we
want to point out that we have not attempted that our models reproduce the
exact values of the observational data points in the millimeter range, due
to the problems discussed in \S \ref {G29dB} and because errors are not
available for some of these data points.

Our best fit model predicts that the flux density at 3.9~$\mu$m should be
lower than the value observed.  We think that the excess of flux density
observed at wavelengths shorter than 3.9~$\mu$m could be due to scattered
light in a cavity carved out by the outflow associated with G29.96$-$0.02
HMPO (Gibb et al. 2004). In fact, Gibb et al. (2004) suggest that
scattered light in G29.96$-$0.02 HMPO could be dominant even at 18 $\mu$m.
However, our models show that the emission from 4-18~$\mu$m can be
reproduced as dust emission from an infalling flattened envelope, and that
scattered light may only be dominant for wavelengths shorter than
$\sim$4~$\mu$m.

 From Table 6 we see that the ratio ${\dot M}/{\dot M_{crit}} = 22,000$. Thus,
we do not expect the development of an \ion{H}{2} region in the G29 HMPO.
This is consistent with the lack of strong free-free emission at the
position of the source.

In order to illustrate the relevance of the millimeter constraints, in
Figure \ref{fig6} we show an additional model (dotted-line), obtained by
fitting only the mid-infrared observations. For this model, we obtain a
value of the luminosity of only 2400 $L_{\odot}$, a density scale of
$5.3\times10^{-12}$ g cm$^{-3}$ and an envelope mass of $\sim$
98~$M_{\odot}$. This set of parameters fits very well the mid-infrared
data; however the model predicts far too little emission at long
wavelengths, underestimating by almost one order of magnitude the
luminosity and mass.  These results illustrate that both millimeter as
well as mid-infrared data are required in order to determine reliable
values of the derived parameters.

It is worthwhile to mention that our models can fit the observed SED with
a value of $R_c\simeq 570$ AU, considerably smaller than the size of the
elongated structures observed towards the G29.96-0.02 HMPO ($\gtrsim$
10,000 AU, Olmi et al. 2003) and some other high-mass protostars
(1000-30000 AU, see Cesaroni 2005 and references therein). These
structures have been interpreted by some authors as tracing large-scale
disks or rotating toroids. Our results suggest that these structures could
be naturally explained as infalling flattened envelopes (with sizes of
thousands of AUs), rather than as large circumstellar disks. The formation
of such disks is expected to occur at a scale of the order of $R_c$, which
according to our models could be of the order of hundreds of AUs.

\subsection{G45.07+0.13 HMPO}

In order to compare the predicted SED of the G45.07+0.13 HMPO with the
data, we use the existing millimeter observations (taken from Su et al.
2004) only as upper limits, since they do not have an angular resolution
high enough to separate the HMPO emission from that of the UC \ion{H}{2}
region. Therefore, we will restrict our comparison to our mid-infrared
observations.

The G45.07+0.13 HMPO exhibits an anomalous spectrum because in spite of
its deep silicate absorption (indicative of a large extinction), it
presents a very strong emission in the near-infrared range (4-8 $\mu$m).
For this source, we also have considered models for both high and low
inclination angles. A model with a highly inclined envelope
($i=58^{\circ}$), a density scale of $\rho_{\rm 1\,AU}= 7.5\times10^{-13}$
g cm$^{-3}$, and a centrifugal radius of $R_c=$270 AU (dotted line in
Figure~\ref{fig7}) predicts too little emission at 4.7 $\mu$m and
shorter wavelengths, and also presents a silicate absorption feature that is too
weak. Alternatively, an envelope with a lower inclination
($i=35^{\circ}$), a larger density scale ($\rho_{\rm 1\,AU}=
5.3\times10^{-12}$ g cm$^{-3}$), and a centrifugal radius of $R_c=$370 AU
(solid line in Figure~\ref{fig7} and Table 5) reproduces better the
emission at 4.7 $\mu$m as well as the depth of the silicate absorption,
although it still has problems in explaining the emission at the shortest
wavelengths ($\sim 2~\mu$m).  Both models require the same luminosity
($L_*=2.5\times 10^4$ $L_{\odot}$), implying a B1 star of 13 $M_{\odot}$.

We favor the low inclination model (solid line in Figure~\ref{fig7} and
Tables 5 and 6) because it can explain reasonably well most of values of
the observed flux density. As we discussed for the G29.96-0.02 HMPO case,
the excess of flux density observed at wavelengths shorter than 3.9~$\mu$m
could be due to scattered light. By running other models with slightly
different values of the parameters, and taking into account the upper
limits set by the submillimeter and millimeter data, we estimate
uncertainties of $\pm 5^\circ$ in the inclination angle, a factor of two
in the density scale and the centrifugal radius, and a factor of three in
the luminosity with respect to the values given in Table 5.

For the values given in Table 6, a ratio $\dot M/\dot M_{\rm crit}=3600$ is
derived. Thus, as in the previous sources, the mass accretion rate for the
G45.07+0.13 HMPO is high enough to quench the development of an UC
\ion{H}{2} region, that is consistent with the lack of detected free-free
emission towards this HMPO.

\section{Summary}

We have obtained mid-infrared data for four HMPO candidates at Gemini
Observatory. These data have the subarcsecond angular resolution necessary
to isolate the emission of the HMPO candidates from that of nearby
sources. The data are well-sampled across the
entire 2 to 25 $\micron$ atmospheric windows, and particularly throughout
the 10 $\micron$ absorption feature.

Since this wavelength range is very sensitive to the source geometry, we
have constructed a grid of SED models for the physical conditions
(density, temperature, luminosity) expected for the early stages of
massive star formation, and taking into account for the first time the
rotation as well as the natural flattening of the infalling envelopes.
These models reach a degree of complexity comparable to those developed
for low-mass protostars. Using our mid-infrared data, as well as
the far-infrared and
millimeter constraints available in the literature, we have fit the
SEDs of the HMPO candidates in G11.94$-$0.62, G29.96$-$0.02, and
G45.07+0.13, inferring physical parameters for the infalling envelopes as
well as for the central stars.

Our main conclusions are the following:

\begin{enumerate}

\item The HMPO candidate in G11.94-0.62 does not seem to be luminous
enough to be considered as a HMPO.  Most likely, it would be better
considered as a ``warm'' core that is forming an intermediate-mass star,
with an infall rate more than one order of magnitude higher than the
typical infall rates of low-mass protostars.

\item The candidates G29.96$-$0.02 and G45.07+0.13 appear to be genuine
HMPOs with high luminosities, $L_* \simeq 20000~L_\odot$, and high
accretion rates, $\dot M \simeq 10^{-3}$-$10^{-2}~M_\odot$~yr$^{-1}$.
These values of the mass accretion rate exceed by several orders of
magnitude the critical value required to quench the development of an UC
\ion{H}{2} region. Therefore, these sources appear to be in a very early
stage, previous to the development of an UC \ion{H}{2} region.

\item Our models are able to fit the observed SEDs of the HMPOs with
values of the centrifugal radius of a few hundred AUs, which are
considerably smaller than the size of the elongated structures observed
towards some high-mass protostars (1000-30000 AU). These structures have
been interpreted by some authors as tracing large-scale disks or rotating
toroids. Our results suggest that these structures could be naturally
explained as infalling flattened envelopes (with sizes of thousands of
AUs), rather than as large circumstellar disks. The formation of such
disks is expected to occur at a scale of the order of $R_c$, which
according to our models could be of the order of hundreds of AUs.

\item Our detailed modeling suggests that massive star formation can
proceed in a way very similar to the formation of low-mass stars.
Unfortunately, massive protostars are more distant than low-mass
protostars, and the instrumentation currently available in the far-infrared and
millimeter ranges does not reach the angular resolution necessary to
separate the HMPO emission from that of other nearby sources. Since these
wavelength ranges are very important to constrain properly the physical
parameters of the sources, such as the luminosity, in general our fit
models are not unique. New data in these wavelength domains would be most
valuable to determine the physical conditions in the early stages of
massive star formation.

\end{enumerate}

\acknowledgments Based on observations obtained at the Gemini Observatory,
which is operated by the Association of Universities for Research in
Astronomy, Inc., under a cooperative agreement with the NSF on behalf of
the Gemini partnership: the National Science Foundation (United States),
the Particle Physics and Astronomy Research Council (United Kingdom), the
National Research Council (Canada), CONICYT (Chile), the Australian
Research Council (Australia), CNPq (Brazil) and CONICET (Argentina).
Gemini program identification numbers associated with the data presented
here are GS-2003B-DD-7 and GS-2004A-Q-7.
M. O. acknowledges support from IAU Peter Gruber Foundation, Junta de
Andaluc\'\i a, MCYT (AYA2002-00376, including FEDER funds), and AECI.
We thank G. Anglada for useful discussions and helpful suggestions.
We also thank F. Wyrowski for communicating us his 1.4~mm data on
G29.96$-$0.02 before publication. Finally, we would also like to thank our
referee for his/her thoughtful review of
our manuscript and useful comments.

Facilities: \facility{Gemini South  (T-ReCS)}.

%% copy editing.  Individual instruments can be provided in parentheses,
%% after the keyword, but they will not be verified.

\clearpage

\begin{table}
\begin{center}
\scriptsize
\caption{T-ReCS Filter Information and Integration and Calibration
Parameters}
\vspace{4mm}
\begin{tabular}{cc|ccc|ccc|cccc}
\hline
\multicolumn{2}{c|}{Filter} &\multicolumn{3}{|c|}{G11.94-0.62}
&\multicolumn{3}{|c|}{G29.96-0.02} &\multicolumn{3}{|c}{G45.07+0.13} \\
%\cline{3-5}\cline{6-8}\cline{9-11}\cline{12-14}
$\lambda_c$    &$\Delta\lambda$  & t & Standard & F$_\nu$ & t & Standard &
F$_\nu$ & t & Standard & F$_\nu$  \\
\hline
2.2  &2.0-2.4   &304 &HD169916    &447     &521 &HD196171    &283    &130 &HD187642   &550     \\
3.9  &3.6-4.1   &304 &HD178345    & 67.0   &521 &HD196171    &122    &130 &HD187642   &200    \\
4.7  &4.4-5.0   &217 &HD178345    & 40.3   &521 &HD196171    & 73.5  &130 &HD187642   &137      \\
7.7  &7.4-8.1   &217 &HD178345    & 17.8   &304 &HD175775    & 29.6  &130 &HD169916   & 47.7  \\
8.7  &8.4-9.1   &217 &HD169916    & 38.1   &521 &HD169916    & 38.1  &130 &HD169916   & 38.1  \\
9.7  &9.2-10.2  &217 &HD169916    & 32.3   &304 &HD175775    & 20.1  &130 &HD169916   & 32.3   \\
10.4 &9.9-10.9  &217 &HD169916    & 28.5   &304 &HD175775    & 17.7  &130 &HD169916   & 28.5   \\
11.7 &11.1-12.2 &217 &HD169916    & 22.4   &304 &HD169916    & 22.4  &130 &HD169916   & 22.4  \\
12.3 &11.7-12.9 &217 &HD169916    & 20.0   &130 &HD169916    & 20.0  &130 &HD169916   & 20.0  \\
18.3 &17.6-19.1 &217 &HD178345    &  3.5   &    &$^a$        &       &130 &HD169916   &  9.6  \\
24.6 &23.6-25.5 &304 &Alpha Cen   & 27.6   &    &\nodata     &       &130 &HD187642   &  5.5   \\
\hline
\end{tabular}
\medskip \\
\end{center}
\scriptsize
 {\em Note:} The value `$\lambda_c$' is the effective central wavelength
of the filter in $\mu$m, and `$\Delta\lambda$' is the wavelength range of
the filter using the 50\% transmission cut-on and cut-off of the bandpass.
The value `t' is the on-source integration time in seconds in the given
filter of the given science field. `Standard' is the name of the standard
star used in the given filter for the given science field. `F$_{\nu}$'
is the assumed flux density in Jy of the standard star in the given
filter. \\
 {\em $^a$} OSCIR data. See De Buizer et al. (2002). \\

\end{table}

\clearpage

\begin{table}
\begin{center}
\scriptsize
\caption{Observed flux densities in mJy for sources on G11.94-0.62 field}
\vspace{4mm}
\begin{tabular}{lcccccc}
\hline
%                    &        &       &Integrated Flux Density (mJy) &    &    & \\
%\cline{2-7}
\hspace{3mm}{$\lambda$$^a$}  & \textbf{DRT03 1 (HMPO)}$^{b}$ & DRT03 2$^{c,d}$ &
DRT03 3$^{c,e}$ & DRT03 4$^{c,f}$ & DRT03 5$^{c,g}$ & DRT03 6$^{h}$\\
\hline
$3.9{\micron} $    & \textbf{15$\pm$2}        &   40$\pm$5     &   71$\pm$8     &   34$\pm$4     &   29$\pm$3     &  52$\pm$7     \\
$4.7{\micron} $    & \textbf{37$\pm$2}        &   87$\pm$6     &  134$\pm$8     &  106$\pm$5     &   85$\pm$3     & 107$\pm$9     \\
$7.7{\micron} $    &\textbf{123$\pm$44}       &  858$\pm$308   & 2250$\pm$808   & 1240$\pm$446   &  854$\pm$308   &2420$\pm$871   \\
$8.7{\micron} $    & \textbf{59$\pm$5}        &  505$\pm$43    &  669$\pm$57    &  523$\pm$45    &  433$\pm$37    & 564$\pm$48    \\
$9.7{\micron} $    & \textbf{14$\pm$3}        &  170$\pm$20    &  118$\pm$17    &  120$\pm$15    &   99$\pm$11    & 113$\pm$18    \\
$10.4{\micron}$    & \textbf{32$\pm$4}        &  412$\pm$51    &  240$\pm$30    &  335$\pm$42    &  296$\pm$37    & 277$\pm$35    \\
$11.7{\micron}$    & \textbf{94$\pm$8}        & 1210$\pm$104   & 1010$\pm$87    & 1150$\pm$99    &  988$\pm$85    & 816$\pm$71    \\
$12.3{\micron}$    & \textbf{80$\pm$19}       & 1350$\pm$316   & 1340$\pm$316   & 1300$\pm$305   & 1050$\pm$247   &1100$\pm$258   \\
$18.3{\micron}$    &\textbf{154$\pm$31}       &11300$\pm$2000  & 8800$\pm$1570  &11600$\pm$2070  & 7000$\pm$1250  &8230$\pm$1470  \\
$20.8{\micron}$$^i$&\textbf{829$\pm$340}  & 5430$\pm$800   & 7820$\pm$1200
&11300$\pm$1640  &11100$\pm$1530  &  \nodata$^j$  \\
$24.6{\micron}$    &\textbf{296$\pm$79}   &35800$\pm$3200  &33600$\pm$3040
&53500$\pm$4800  &28700$\pm$2590  &  \nodata$^j$  \\
\hline
\end{tabular}
\medskip \\
\end{center}
\scriptsize
 {\em Note:} Errors given to the flux values are 1-$\sigma$ errors that
are quadrature addition of the photometric and calibration errors. \\
 {\em $^a$} Effective central filter wavelength. \\
 {\em $^b$} Integrated flux density in a 0$\farcs$6 radius aperture
centered on the HMPO candidate. \\
 {\em $^c$} This is an extended emission source and is not fully resolved
from other sources on the field. \\
 {\em $^d$} Integrated flux density in a 2$\farcs$1 radius aperture
centered on the center of the extended mid-infrared emission from this source.\\
 {\em $^e$} Integrated flux density in a 2$\farcs$7 radius aperture
centered on the center of the extended mid-infrared emission from this source.\\
 {\em $^f$} Integrated flux density in a 1$\farcs$8 radius aperture
centered on the center of the extended mid-infrared emission from this source.\\
 {\em $^g$} Integrated flux density in a 1$\farcs$2 radius aperture
centered on the center of the extended mid-infrared emission from this source.\\
 {\em $^h$} This single extended source was labeled as two individual
sources in the work of De Buizer et al. (2003, 2005), however the
observations presented here show it to be a very large and amorphous
single but extended source. The integrated flux density is from a
3$\farcs$1 radius aperture centered on the center of the extended
mid-infrared emission from this source.\\
 {\em $^i$} From MIRLIN/IRTF data used in De Buizer et al. (2003, 2005). \\
 {\em $^j$} Source partially off field at this wavelength so no flux
density given. \\

 \end{table}

 \clearpage

\begin{table}
\begin{center}
\scriptsize
\caption{Observed flux densities in mJy for sources on G29.96-0.02 field}
\vspace{4mm}
\begin{tabular}{lccc}
\hline
%                    &               & Integrated Flux Density (mJy) &    \\
%\cline{2-4}
%\hspace{3mm}\raisebox{1.5ex}[0pt]{$\lambda$$^a$}
\hspace{3mm}{$\lambda$$^a$}& \textbf{HMPO}$^b$& HMPO($+$bg)$^c$ & UC HII$^d$
\\
\hline
$2.2{\micron}$       &\textbf{$<$5}           &14   $\pm$2      &327$\pm$24        \\
$3.9{\micron}$       &\textbf{38  $\pm$5}     &112  $\pm$10     &4300$\pm$368      \\
$4.7{\micron}$       &\textbf{123 $\pm$12}    &321  $\pm$6      &10800$\pm$183     \\
$7.7{\micron}$       &\textbf{66  $\pm$20}    &492  $\pm$85     &37400$\pm$6440    \\
$8.7{\micron}$       &\textbf{13  $\pm$2}     &437  $\pm$31     &77800$\pm$5550    \\
$9.7{\micron}$       &\textbf{$<$25}          &239  $\pm$26     &56000$\pm$6090    \\
$10.4{\micron}$      &\textbf{13  $\pm$3}     &370  $\pm$45     &80000$\pm$9700    \\
$11.7{\micron}$      &\textbf{197 $\pm$32}    &1130 $\pm$86     &141000$\pm$10800  \\
$12.3{\micron}$      &\textbf{345 $\pm$78}    &1680 $\pm$342    &169000$\pm$34000  \\
$18.3{\micron}^{e}$  &\textbf{2280$\pm$340}   &16400$\pm$1640   &462000$\pm$46000  \\
\hline
\end{tabular}
\medskip \\
\end{center}
\scriptsize
 {\em Note:} Values preceded by a `$<$' means it is a 3-$\sigma$ upper
limit of the source flux within the apertures specified in the table
notes below. Errors given to the flux values are 1-$\sigma$ errors that
are
quadrature addition of the photometric and calibration errors. \\
 {\em $^a$} Effective central filter wavelength. \\
 {\em $^b$} Integrated flux density after subtraction of background due to
extended flux from the UC \ion{H}{2} region. \\
 {\em $^c$} Integrated flux density in a 1$\farcs$25 radius aperture
centered on the HMPO peak without subtraction of the background UC \ion{H}{2}
region.\\
 {\em $^d$} Integrated flux density in a 6$\farcs$4 radius aperture
centered on the center of the extended mid-infrared emission from the UC \ion{H}{2} region.\\
 {\em $^e$} From OSCIR data used in De Buizer et al. (2002). \\

\end{table}

 \clearpage

\begin{table}
\begin{center}
\scriptsize
\caption{Observed flux densities in mJy for sources on G45.07+0.13 field}
\vspace{4mm}
\begin{tabular}{lccccccc}
\hline
\hspace{3mm}{$\lambda$$^a$}       & DRT03 1$^b$ & DRT03 2$^c$ & \textbf{DRT03 3
(HMPO)}$^d$ & DOC05 4$^e$ & DOC05 5$^f$ & DOC05 6$^g$ & DOC05 7$^g$\\
\hline
$2.2{\micron}$      &$<$14          &    17$\pm$2         &   \textbf{15$\pm$2}    &$<$4.04      &   8$\pm$2   &$<$4.04     &$<$4.04      \\
$3.9{\micron}$      &62   $\pm$7    &  1460$\pm$156       & \textbf{1440$\pm$153}  &$<$3.1       & 567$\pm$60  &$<$3.1      &$<$3.1       \\
$4.7{\micron}$      &153  $\pm$10   &  4950$\pm$81        & \textbf{4880$\pm$80}   &$<$8.3       &1610$\pm$27  &$<$8.3      &$<$8.3       \\
$7.7{\micron}$      &1330 $\pm$331  & 13700$\pm$3390      & \textbf{7750$\pm$1910} &  72$\pm$20  &2710$\pm$670 & 25$\pm$12  &  71$\pm$20  \\
$8.7{\micron}$      &475  $\pm$38   &  5940$\pm$467       &  \textbf{841$\pm$66}   &  24$\pm$3   & 996$\pm$78  &  9$\pm$2   &  14$\pm$2   \\
$9.7{\micron}$      &233  $\pm$27   &  2080$\pm$197       &  \textbf{165$\pm$17}   &$<$14.6      & 308$\pm$30  &$<$14.6     &$<$14.6      \\
$10.4{\micron}$     &377  $\pm$41   &  4350$\pm$461       &  \textbf{472$\pm$50}   &17  $\pm$4   & 692$\pm$73  &  9$\pm$3   &$<$9.4       \\
$11.7{\micron}$     &1260 $\pm$88   & 18100$\pm$1270      & \textbf{4720$\pm$330}  &66  $\pm$5   &2350$\pm$164 & 26$\pm$3   &  19$\pm$3   \\
$12.3{\micron}$     &2310 $\pm$347  & 37100$\pm$5560      &\textbf{11200$\pm$1670} &109 $\pm$17  &6610$\pm$991 & 48$\pm$10  &  61$\pm$11  \\
$18.3{\micron}$     &10900$\pm$964  & 72900$\pm$6360      &\textbf{12000$\pm$1050} &544 $\pm$66  &8600$\pm$753 &$<$136      & 264$\pm$51  \\
$20.8{\micron}^{h}$ &15100$\pm$2130 &102000$^{i}\pm$13500 &\textbf{19700$\pm$2780} &\nodata      &\nodata      &\nodata     &\nodata      \\
$24.6{\micron}$     &51700$\pm$4150 &564000$\pm$45000     &\textbf{67600$\pm$5400} &2170$\pm$208 &\nodata      &335$\pm$119 &1338$\pm$159 \\
\hline
\end{tabular}
\medskip \\
\end{center}
\scriptsize
 {\em Note:} Values preceded by a `$<$' means it is a 3-$\sigma$ upper
limit of the source flux within the apertures specified in the table
notes below. Errors given to the flux values are 1-$\sigma$ errors that
are
quadrature addition of the photometric and calibration errors. \\
 {\em $^a$} Effective central filter wavelength. \\
 {\em $^b$} Integrated flux density in a 2$\farcs$4 radius aperture
centered on center of the extended mid-infrared emission from this source.
\\
 {\em $^c$} Integrated flux density in a 2$\farcs$0 radius aperture
centered on the peak of this source. This source flux density is
contaminated by some emission from DOC05 5, which is a close mid-infrared
companion. DOC05 5 was fit by a gaussian and subtracted off before the
aperture photometry was performed on DRT03 2. However DOC05 5 is much
fainter than DRT03 2, so the contamination is relatively small (on order
the flux error) at all wavelengths. \\
 {\em $^d$} Integrated flux density in a 1$\farcs$1 radius aperture
centered on the peak of this HMPO candidate.\\
 {\em $^e$} Integrated flux density in a 0$\farcs$7 radius aperture
centered on the peak of this source.\\
 {\em $^f$} Flux densities quoted for this source are rough estimates.
DOC05 5 is faint and separated by only 0$\farcs$7 from the very prominent
mid-infrared source DRT03 2. Fluxes for this source were obtained by
fitting DRT03 2 by a gaussian and subtracting it away before performing
aperture photometry with a 0$\farcs$7 radius aperture. However, as the
source was observed at longer and longer wavelengths resolution grew
steadily worse until at 24.5$\micron$ the sources could not be resolved.\\
 {\em $^g$} Integrated flux density in a 0$\farcs$7 radius aperture
centered on the peak of this source.\\
 {\em $^h$} From MIRLIN/IRTF data used in De Buizer et al. (2003, 2005). \\
 {\em $^i$} This is the combined flux from DRT03 2 and DOC05 5, since they
are not resolved from each other using IRTF. \\

\end{table}

\clearpage

\begin{table}
\begin{center}
\caption{Parameters of the Best Fit Models}\label{tbl5}
\begin{tabular}{lccccccc}
\hline
HMPO &$\eta$ &$R_{\rm in}$ &$R_{\rm out}$ &$R_{\rm c}$ &$L_*$ &$\rho_{\rm
1\,AU}$ &$i$
\\
&&(AU) &(AU) &(AU) &($L_{\odot}$) &(g cm$^{-3}$) &($^\circ$)   \\
\hline
G11.94-0.62 &2.5 &2   &5000  &30  &75    &1.5$\times10^{-13}$ &53 \\
G29.96-0.02 &2.5 &245 &12000 &570 &18000    &3.0$\times10^{-11}$ &12 \\
G45.07+0.13 &2.5 &227 &9000  &370 &25000 &5.3$\times10^{-12}$ &35 \\
\hline
\end{tabular}
\end{center}

\end{table}

\clearpage

\begin{table}
\begin{center}
\caption{Physical Parameters Derived from the Models\label{tbl6} }
\begin{tabular}{lccccccc}
\hline
HMPO &$M_{\rm env}$\tablenotemark{a} &$T_{\rm 1000\,AU}$\tablenotemark{b}
&$M_*$\tablenotemark{c} &Spectral\tablenotemark{d} &$\dot
M$\tablenotemark{e} &$\dot N_i$\tablenotemark{f} &$\dot M_{\rm
crit}$\tablenotemark{g}\\
&($M_{\odot}$) &(K) &($M_{\odot}$) &Type &($M_{\odot}$ yr$^{-1}$)
&(s$^{-1}$) &($M_{\odot}$ yr$^{-1}$)\\
\hline
G11.94-0.62  &1  &40 &5 &A0 &$8.7\times 10^{-5}$ &$3.0\times 10^{37}$
&$3.5\times 10^{-11}$\\
G29.96-0.02  &576 &300 &11 &B1 &$2.0\times 10^{-2}$ &$1.0\times 10^{46}$
&$9.0\times 10^{-7}$ \\
G45.07+0.13  &64 &200 &13 &B1 &$3.6\times 10^{-3}$ &$1.0\times 10^{46}$
&$1.0\times 10^{-6}$\\
\hline
\end{tabular}
\end{center}
\tablenotetext{a}{Mass of the envelope, obtained by integration of the
density distribution}
\tablenotetext{b}{Temperature of the envelope at a radius of 1000~AU.}
\tablenotetext{c}{Mass of the central star, obtained from $L_*$ using the
evolutionary tracks of Behrend \& Maeder 2001.}
\tablenotetext{d}{Obtained from tables of Doyon (1990) using the
luminosities given in Table 5.}
\tablenotetext{e}{Mass accretion rate, obtained from $\rho_{\rm 1\,AU}$
and $M_*$.}
\tablenotetext{f}{Rate of ionizing photons, obtained from  Doyon (1990).}
\tablenotetext{g}{Critical mass accretion rate,
$\dot M_{\rm crit}={\left(8 \pi G m^2_{\rm H} \alpha^{-1} \dot N_i
M_*\right)}^{1/2}$, where $m_{\rm H}$ is the hydrogen mass, and
$\alpha$ is the recombination coefficient ($n>1$) obtained from Yorke
1984, and Walmsley 1995 assuming spherical accretion.}

\end{table}

\clearpage

\clearpage

\begin{figure}
\epsscale{0.81}
\plotone{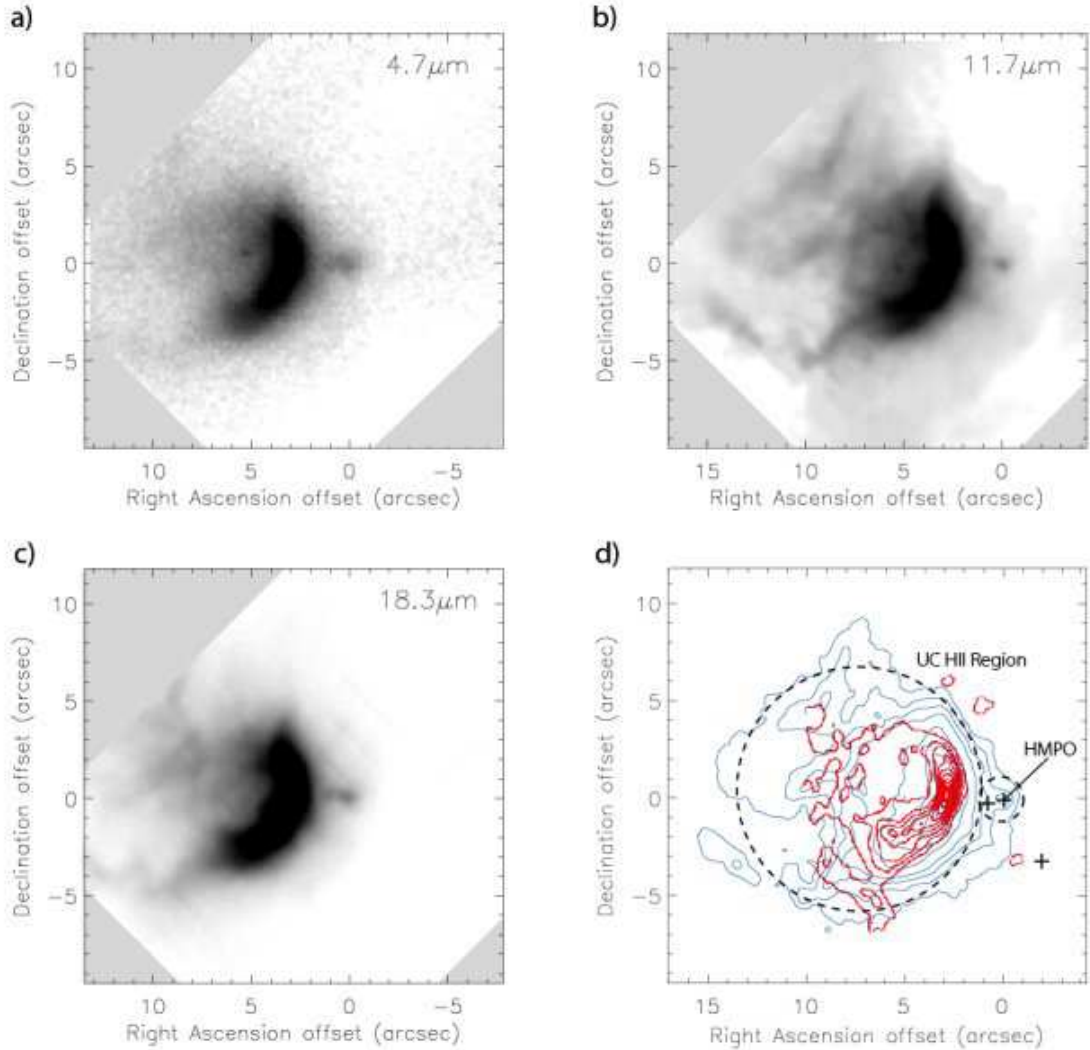}
 \caption{Images of the G29.96-0.02 field taken at a) 4.7 $\mu$m, b)
11.7 $\mu$m, and c) 18.3 $\mu$m at Gemini South with T-ReCS. These images
had to be de-rotated to show north up and east left, hence the gray areas
with no data in the corners of the panels.  d) A contour plot (smoothed by
0$\farcs$18) of the 11.7 $\mu$m image (blue contours) of the field
showing sources with their labels and encircled with dashed lines that
represent the apertures used for photometry. These aperture radii are
given in the notes in Table 3.  Plotted as crosses are the locations
of the water maser groups as given by Hofner \& Churchwell (1996). Also
plotted (red contours) are the 2 cm radio continuum contours from Hofner
\& Churchwell (1996). The origin of each panel is the mid-infrared peak of
the HMPO candidate (R.A.(J2000) = 18$^h$ 46$^m$ $03\rlap.^s74$,
Decl.(J2000) = $-02^\circ$ $39'$ $22\farcs1$).
 \label{fig2}}
\end{figure}

\clearpage

\begin{figure}
\epsscale{0.81}
\plotone{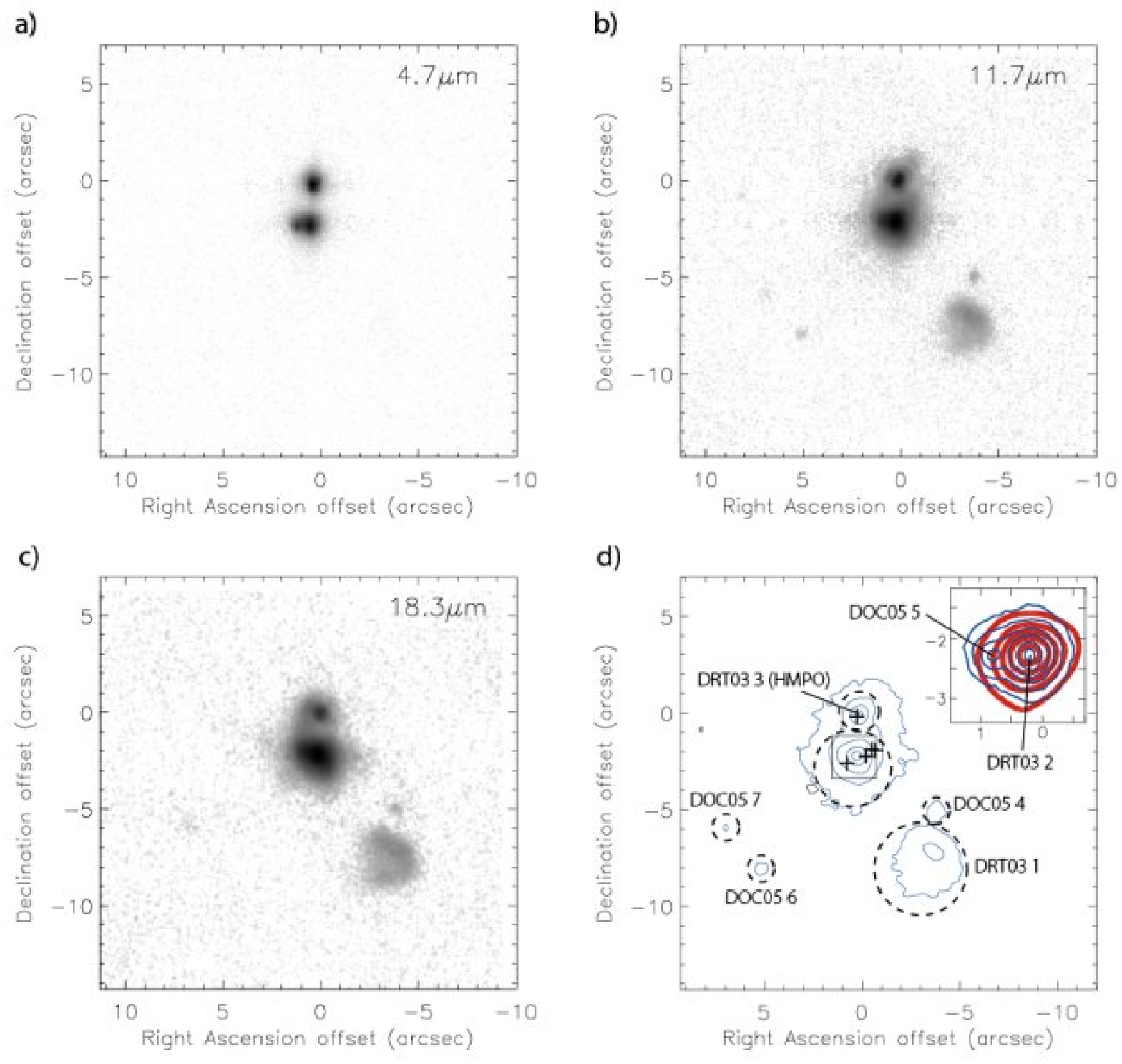}
 \caption{Images of the G45.07+0.13 field taken at a) 4.7 $\mu$m, b) 11.7
$\mu$m, and c) 18.3 $\mu$m at Gemini South with T-ReCS. d) A contour plot
(smoothed by 0$\farcs$27) of the 11.7 $\mu$m image (blue contours) of the
field showing sources with their labels and encircled with dashed lines
that represent the apertures used for photometry.  These aperture radii
are given in the notes in Table 4. In the upper-right corner there
is a blow-up of the the region encompassed by the square. This blow-up
shows contours of the double source, DRT03 2 and DOC05 5, at 4.7 $\mu$m
and is overlaid (red contours) with the 2 cm radio continuum
contours from Hofner \& Churchwell (1996). Also plotted as crosses are the
locations of the water maser groups as given by Hofner \& Churchwell
(1996). The origin of each panel is the mid-infrared peak of the HMPO
candidate (R.A.(J2000) = 19$^h$ 13$^m$ $22\rlap.^s07$, Decl.(J2000) =
$+10^\circ$ $50'$ $55\farcs4$).
 \label{fig3}}
\end{figure}

\clearpage

\begin{figure}
\plotone{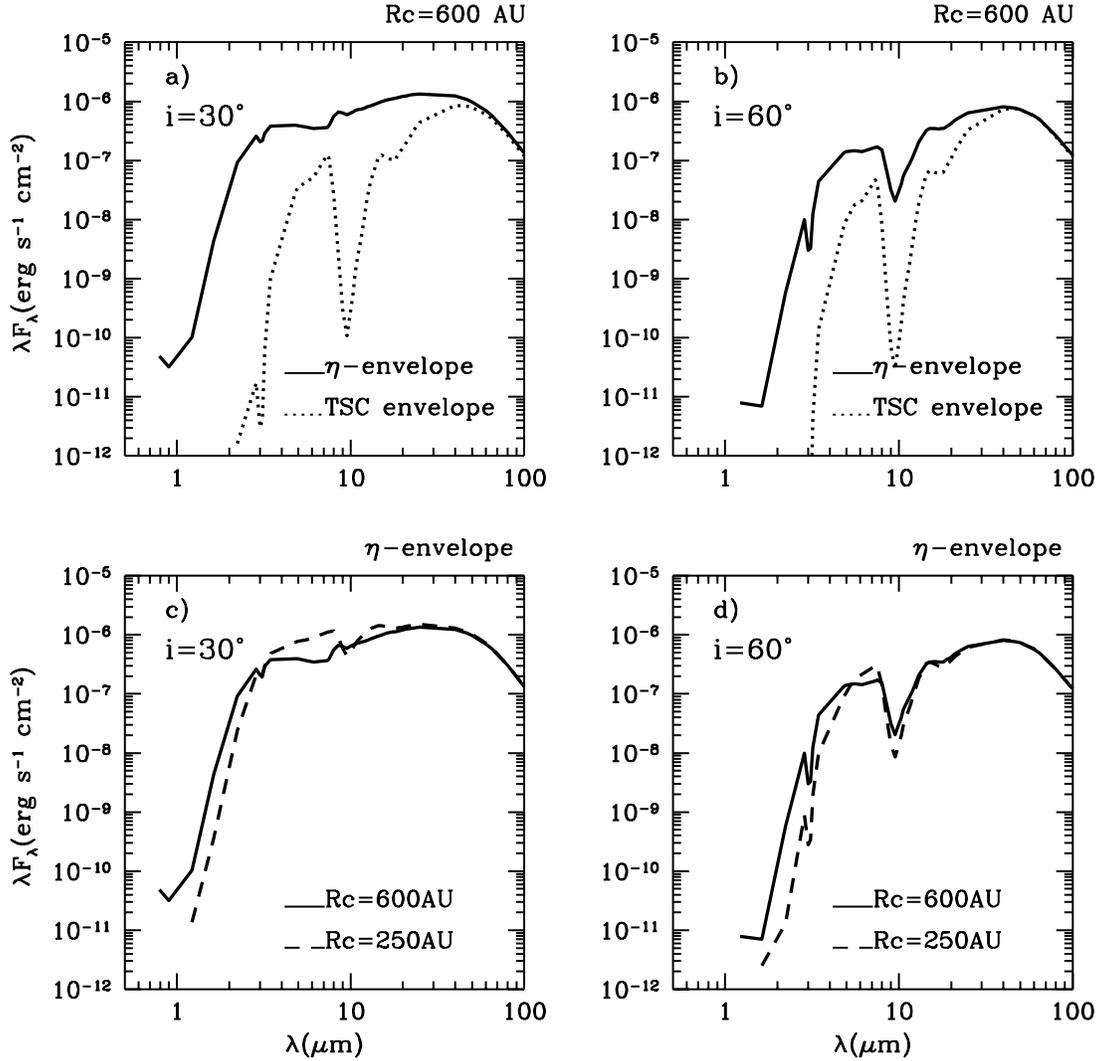}
 \caption{(a) Model SEDs for an $\eta=2$ envelope (solid-line) and a TSC
envelope (dotted-line) with the same values of the centrifugal radius
($R_c=600$ AU) and inclination angle ($i=30^{\circ}$). (b) Same as (a),
but for a higher value of the inclination angle ($i=60^{\circ}$).  (c)
Model SEDs for two $\eta=2$ envelopes, with the same value of the
inclination angle ($i=30^{\circ}$)  and two different values of the
centrifugal radius:  $R_c=250$ AU (dashed line) and $R_c=600$ AU (solid
line). (d) Same as (c), but for $i=60^{\circ}$. The stellar luminosity
($L_*=25000~L_{\odot}$), the mass of the envelope (M$_{env}=9~M_{\odot}$),
and the assumed distance (1 kpc) are the same in all models.}
 \label{fig4}
 \end{figure}

\begin{figure}
\plotone{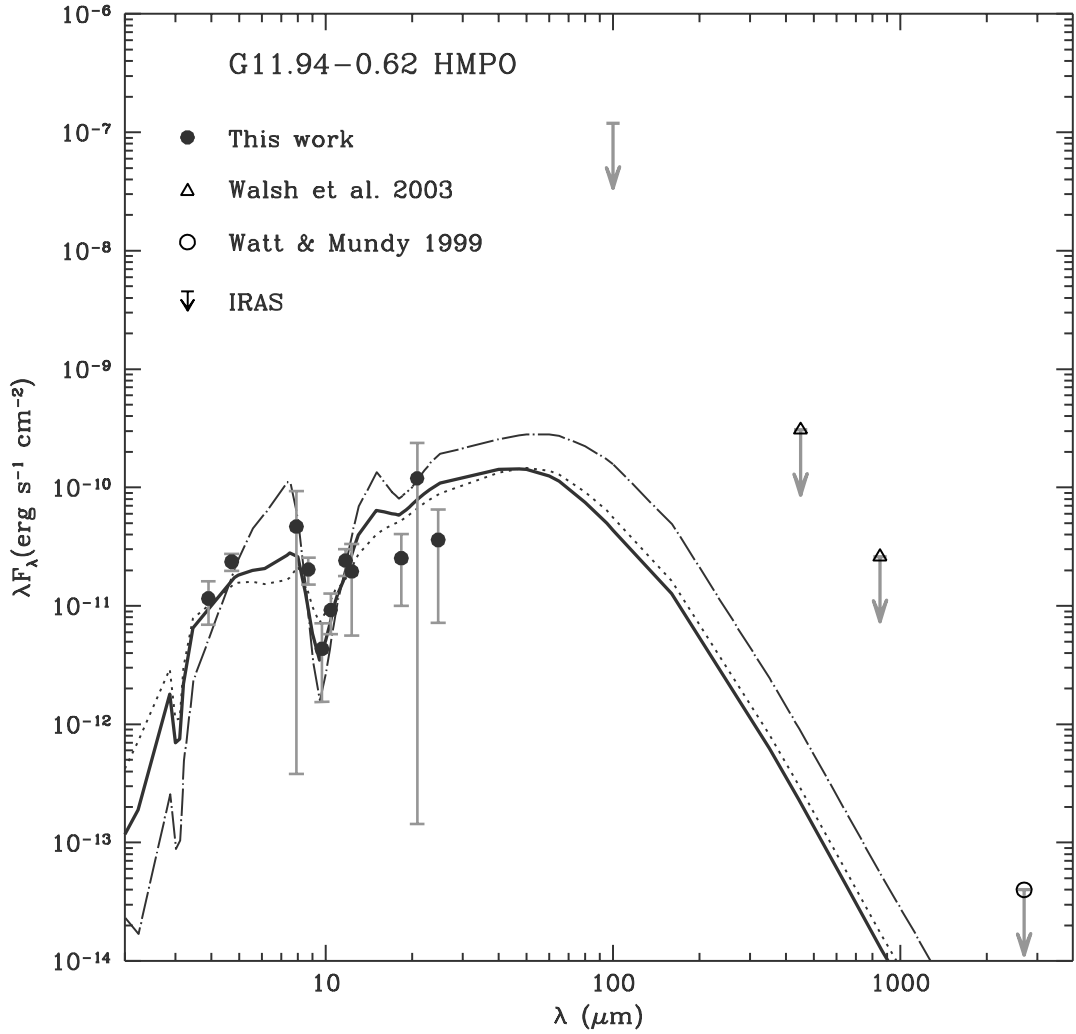}
 \caption{Observed and model SEDs for the source G11.94-0.62 HMPO. The
different symbols represent the observed values of the flux density. Error
bars are 3-$\sigma$. Upper limits are represented by arrows. Solid-line
represents the best fit model, obtained with $i=53^{\circ}$, $R_c=30$ AU,
$L_*=75~L_{\odot}$, $\rho_{\rm 1\,AU}=2 \times 10^{-13}$ g cm$^{-3}$ (see
Table 5), while the dotted line represents a model with the same $i$ and
$\rho_{\rm 1\,AU}$ but with $R_c=100$ AU. Dot-dashed line represents a
model with a low inclination angle of $i=30^{\circ}$, with $R_c=30$ AU and
$\rho_{\rm 1\,AU}=7.5 \times 10^{-13}$ g cm$^{-3}$. The adopted distance
is 4.2 kpc (Hofner \& Churchwell 1996). }
 \label{fig5}
 \end{figure}

\begin{figure}
\plotone{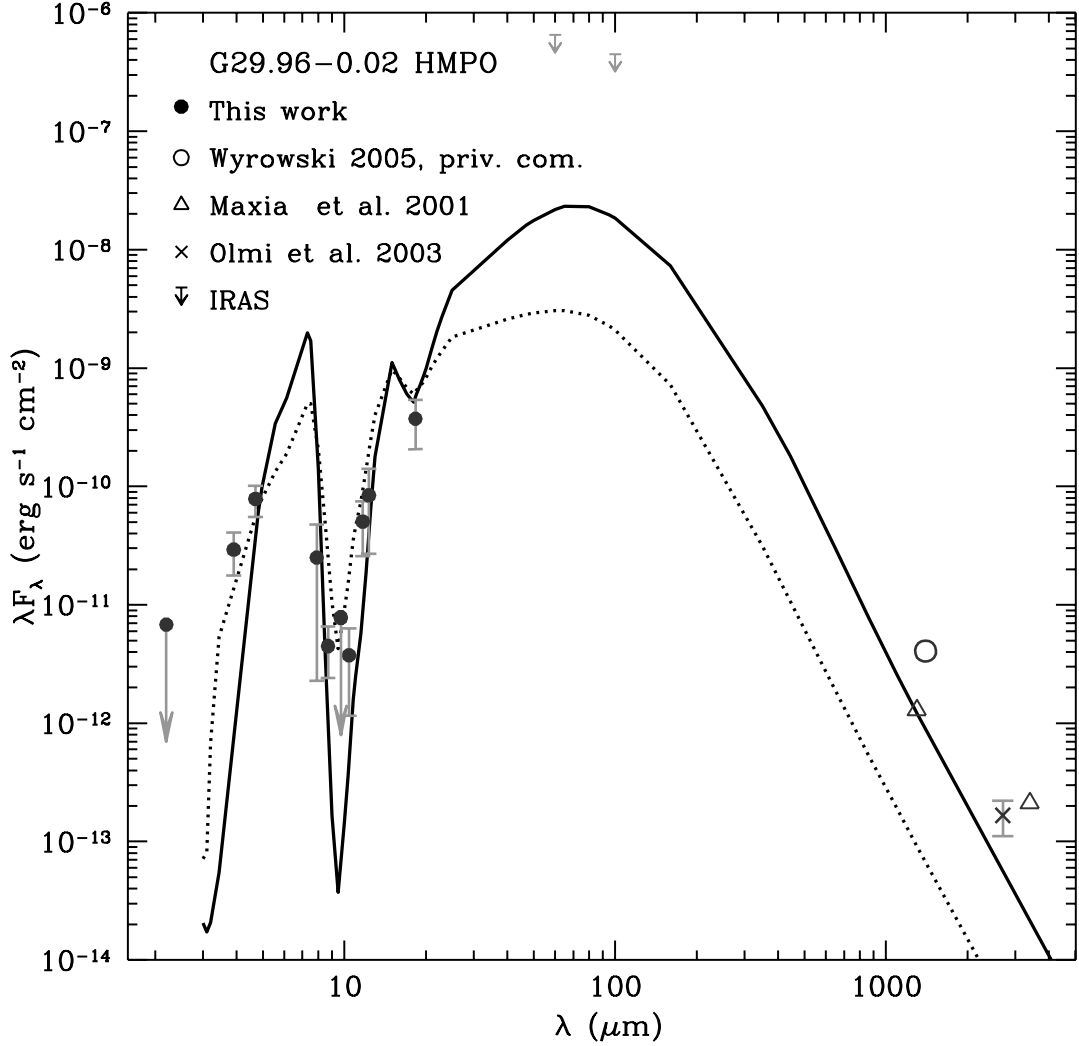}
 \caption{Observed and model SEDs for the source G29.96-0.02 HMPO.  The
different symbols represent the observed values of the flux density. Error
bars are 3-$\sigma$ for the IR data. The 2.7 mm error bars correspond to
the uncertainties assigned by Olmi et al. 2003 due to the subtraction
method employed. No errors are available for the remaining millimeter flux
densities. Upper limits are represented by arrows. Solid line represents
the best fit model, obtained with $i=12^{\circ}$, $R_c=570$ AU,
$L_*=1.8\times 10^{4}~L_{\odot}$, and $\rho_{\rm 1\,AU}=3 \times 10^{-11}$
g cm$^{-3}$ (see Table 5). Dotted line represents the best fit model
obtained by using only the mid-infrared data, with $i=12^{\circ}$,
$R_c=380$ AU, $L_*=2.4\times 10^{3}~L_{\odot}$, and $\rho_{\rm 1\,AU}=5
\times 10^{-12}$ g cm$^{-3}$. Note that the 9.7~$\mu$m is only an upper
limit and cannot discriminate between the two models in terms of the depth
of the absorption feature. The adopted distance is 8.4 kpc (Sewilo et al.
2004). Note that both models have the same inclination angle, but differ
significantly in the luminosity and density. }
 \label{fig6}
 \end{figure}

\begin{figure}
\plotone{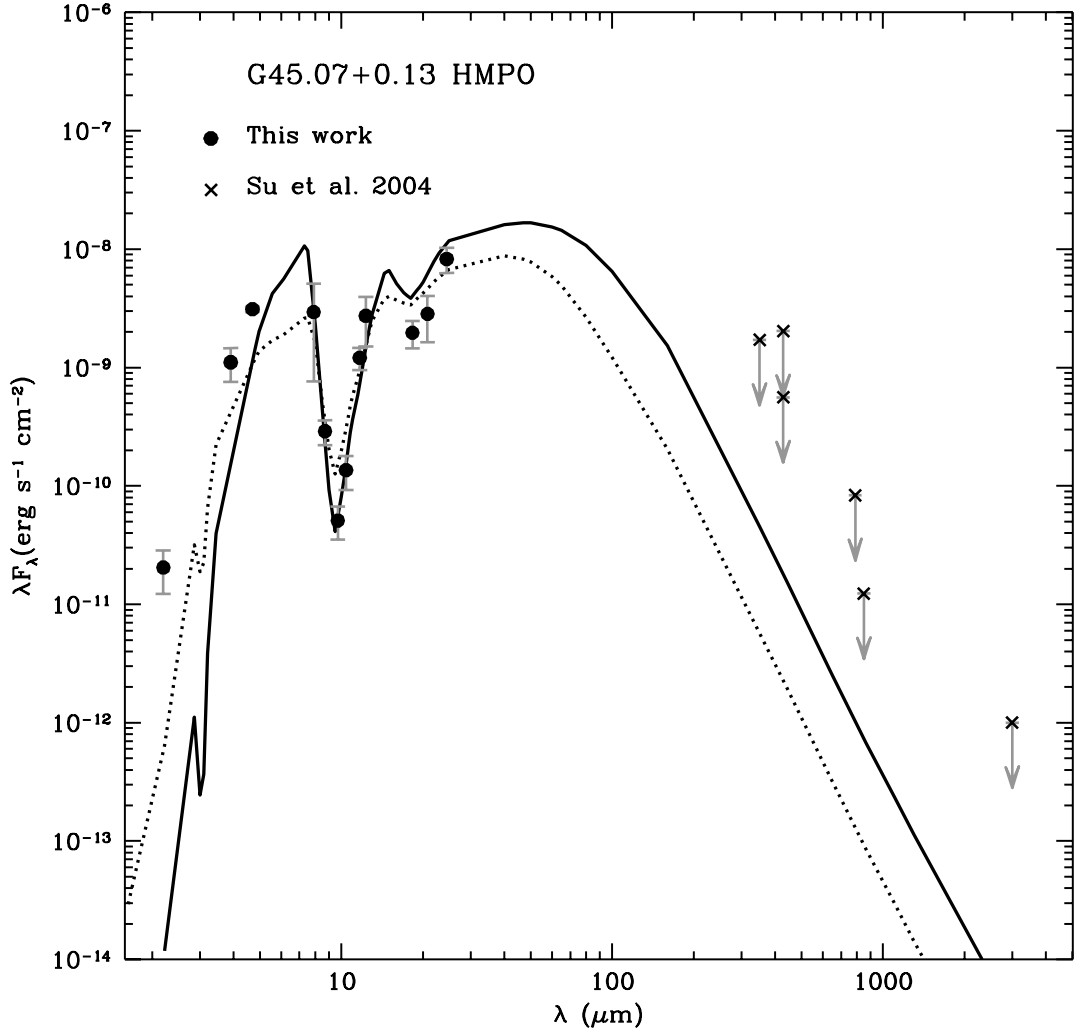}
 \caption{Observed and model SEDs for the source G45.07+0.13 HMPO.  The
different symbols represent the observed values of the flux density. Error
bars are 3-$\sigma$. Upper limits are represented by arrows. Solid line
represents the best fit model, obtained with $i=30^{\circ}$, $R_c=370$ AU,
$L_*=2.5\times 10^{4}~L_{\odot}$, and $\rho_{\rm 1\,AU}=5.3 \times
10^{-12}$ g cm$^{-3}$ (see Table 5). Dotted line represents a model with
the same luminosity and a higher inclination angle, $i=58^{\circ}$, with
$R_c=270$ AU, and $\rho_{\rm 1\,AU}=7.5 \times 10^{-13}$ g cm$^{-3}$. The
adopted distance is 9.7 kpc (Wood \& Churchwell 1989).}
 \label{fig7}
 \end{figure}

\end{document}